\documentclass[sn-mathphys]{sn-jnl}

\usepackage[utf8]{inputenc}
\usepackage[T1]{fontenc}
\usepackage{mathtools}
\usepackage{bm}                  
\usepackage{standalone}
\usepackage{multicol}        
\usepackage{multirow}
\usepackage{siunitx}
\usepackage{subcaption}
\usepackage{shellesc}
\usepackage{wrapfig}
\usepackage{amssymb}
\usepackage{amsthm}
\usepackage{amsmath}
\usepackage{xspace}
\usepackage[super]{nth}
\usepackage{scalerel,stackengine}
\usepackage{natbib}
\usepackage{siunitx}
\usepackage{hyperref}

\hypersetup{
	pdftitle={A Time-Accurate Inflow Coupling for Zonal LES},
	pdfauthor={Marcel Blind; Johannes Kleinert; Andrea Beck; Thorsten Lutz},
}

\newcommand{\fref}[1]{Fig.~\ref{#1}}
\newcommand{\tref}[1]{Tab.~\ref{#1}}
\newcommand{\sref}[1]{Sec.~\ref{#1}}
\newcommand*{\eg}{e.g.\@\xspace}

\newcommand*{\cf}{cf.\@\xspace}

\newcommand{\norm}[1]{\left\lVert#1\right\rVert_{L_2}}

\jyear{2022}
\begin{document}

\title[A Time-Accurate Inflow Coupling for Zonal LES]{A Time-Accurate Inflow Coupling for Zonal LES}

\author*[1]{\fnm{Marcel} \sur{Blind}}\email{blind@iag.uni-stuttgart.de}

\author[1]{\fnm{Johannes} \sur{Kleinert}}\email{kleinert@iag.uni-stuttgart.de}

\author[1]{\fnm{Thorsten} \sur{Lutz}}\email{lutz@iag.uni-stuttgart.de}

\author[2,1]{\fnm{Andrea} \sur{Beck}}\email{beck@iag.uni-stuttgart.de}

\affil[1]{\orgdiv{Institute of Aerodynamics and Gas Dynamics}, \orgname{University of Stuttgart}, \orgaddress{\street{Pfaffenwaldring 21}, \city{Stuttgart}, \postcode{70569}, \country{Germany}}}

\affil[2]{\orgdiv{Institute of Fluid Dynamics and Thermodynamics}, \orgname{Otto-von-Guericke-University Magdeburg}, \orgaddress{\street{Universitätsplatz 2}, \city{Magdeburg}, \postcode{39106}, \country{Germany}}}

\abstract{Generating turbulent inflow data is a challenging task in zonal Large Eddy Simulation (zLES) and often relies on predefined DNS data to generate synthetic turbulence with the correct statistics. The more accurate, but more involved alternative is to use instantaneous data from a precursor simulation. Using instantaneous data as an inflow condition allows to conduct high fidelity simulations of subdomains of \eg an aircraft including all non-stationary or rare events. In this paper we introduce a tool-chain that is capable of interchanging highly resolved spatial and temporal data between flow solvers with different discretization schemes. To accomplish this, we use interpolation algorithms suitable for scattered data in order to interpolate spatially. In time we use one-dimensional interpolation schemes for each degree of freedom. The results show that we can get stable simulations that map all flow features from the source data into a new target domain. Thus, the coupling is capable of mapping arbitrary data distributions and formats into a new domain while also recovering and conserving turbulent structures and scales. The necessary time and space resolution requirements can be defined knowing the resolution requirements of the used numerical scheme in the target domain.}

\keywords{DGSEM, instantaneous inflow condition, coupling, zonal LES}

\maketitle

\section{Introduction}\label{introduction}

In modern Computational Fluid Dynamics (CFD) research Large Eddy Simulation (LES) is becoming more popular due to increased computation performance \cite{Flad:2020}. However, many practical applications still are out of range for a detailed investigation using this technique. Therefore, many simulations run today are based on so called zonal approaches that often depend on predefined DNS data to generate synthetic turbulence with the correct statistics. By zonal we mean that only a small subset of a domain is simulated with the LES method in order to decrease computational cost, while the majority of the domain is for example solved by a much cheaper Reynolds-Averaged Navier-Stokes (RANS) method, or the subset is equipped with suitable boundary conditions, in particular scale-resolving inflow data. This can be of special interest in the simulation of turbulent boundary layers, where we do not want to simulate the initial transition process, but are just interested in the fully developed boundary layer as a starting point. To achieve this, there have been developed many approaches, such as the synthetic eddy method \cite{Jarrin:2006,Jarrin:2009} or the recycling-rescaling approach \cite{Lund:1998,Kuhn:2020} which allow for significantly smaller domains. A practical example is the simulation of acoustic noise at the trailing edge of an airfoil where a detailed simulation of only a part of the airfoil is needed \cite{Kempf:2022}. One similarity of the applications just described is their dependency on boundary layer properties and therefore reference data from literature.

Another slightly different example is the investigation of the interaction of an incoming turbulent wake with the boundary layer. For example this can be applied to the interaction of a turbulent wing wake with the horizontal stabilizer of an aircraft (\cf \fref{fig:tandem_overview}). The described scenario poses a challenge, since fully scale resolving codes often can not afford to compute the whole aircraft and codes that are capable of running a simulation of a whole aircraft are often not able to run high fidelity simulation of parts of it. Therefore, there is the need to map the results in a time-accurate manner from one simulation to an inflow/initial condition on a detailed simulation with a smaller domain and to impose them as inflow conditions. This approach allows for refined simulation of areas of special interest. Also such a coupling between these codes enables a way to further investigate the interaction between turbulence of different physical scales very efficiently. Therefore, zonal simulations of high Reynolds number flow become feasible.

When trying to couple different numerical codes we encounter several problems on how to approach this:
\begin{enumerate}
	\item Is a two-way coupling necessary or is one-way sufficient?
	\item Do we couple the codes during runtime?
	\item Are the underlying numerics compatible?
\end{enumerate}

The first question is - in context of the scenario described above - easy to answer. Since we only are interested in the effects of an incoming flow to the target domain, a one-way coupling is sufficient. We note that, depending on the equation system, a one-way coupling via a prescribed Dirichlet boundary is prone to errors, since information transport is limited to one direction. However, we can justify this simplification by assuming that we \eg extract the flow in a wake region with no proximity to a wall in case of the compressible Navier-Stokes equations. Additionally, the simplification removes a lot of complexity and thus enables efficient coupling of different solvers and experiments which would not be possible in a two-way coupled way.

Thus, we can directly answer the second question. Having both codes run separately allows us to perform the mapping in a preprocessing step for the zonal simulation and thus removes a bottleneck during runtime. In addition, it avoids the complexities of having to solve the High Performance Computing (HPC) problem of having to run two possibly very heterogeneous codes synchronously and establish efficient parallel communication patterns. As mentioned before the coupling is designed to perform detailed simulations of a subdomain, meaning that the area of special interest is also contained in the full domain simulation and therefore is assumed to be represented in a sufficient way to capture all the necessary physics. Also we have to consider that the incoming physics can be truncated. We thus investigate the influence of different resolution combinations in order to quantify this error.

The third question is harder to answer since we not only have to take the spatial discretization like finite volume, finite difference and finite element methods into account, but also take care of the temporal discretization, which in most applications is either implicit or explicit. In general, two choices for mapping the solutions between two heterogeneous representations are possible: A projection approach, and an interpolation method. While the former is (approximately) conservative, ensuring this property on arbitrary meshes in space and time is cumbersome, expensive (as it requires non-local operations) and not very flexible. The interpolation approach relaxes this condition, making the mapping process very general and allows for extending the algorithms to work with arbitrary $(x,y,z)$ data as an input and map it to a compatible data format. The mapping algorithms thus have to be able to capture resolution differences from both grid spacing and numerical efficiency per DOF, arbitrary points and inconsistent time steps. Hence, interpolation algorithms seem to be a good choice for achieving these properties in space and time.

Another problem we have to tackle is how to deal with large data sets. Although we opt for an offline coupling which avoids having to transfer the data in situ, time resolved surface or volume data is very memory intensive. Thus, memory management algorithms have to be taken into account, including parallelization, load balancing and data reduction in order to keep compute costs low.

In this paper we want to show that mapping instantaneous data from the DLR finite volume code TAU to the high-order spectral element code FLEXI is possible and allows for detailed simulation of subdomains. Thus, we want to answer the following research questions:
\begin{enumerate}
	\item Is it possible to get a mapping for the data which allows for a numerically stable coupled simulation?
	\item Which temporal resolution is needed?
\item How does the difference in spatial resolution (mesh and numerical efficiency per DOF) affect the mapping error?
\end{enumerate}
 \section{Numerical Methods}\label{numerics}

An important aspect for the mapping algorithm is the knowledge of the underlying numerics and the associated scale resolving properties which we are going to assess in more detail in the following section.

\subsection{Code Frameworks}

The target numerical scheme for which the data has to be prepared is the open-source discontinuous Galerkin framework FLEXI, developed at the University of Stuttgart \cite{Krais:2021}. In this scheme the domain is partitioned into non-overlapping unstructured hexahedral elements. We can choose an arbitrary polynomial degree for the elements. This also means that the solution in each element is represented as a polynomial. 

\begin{figure}[!htb]
	\centering
	\includegraphics[width=0.85\columnwidth]{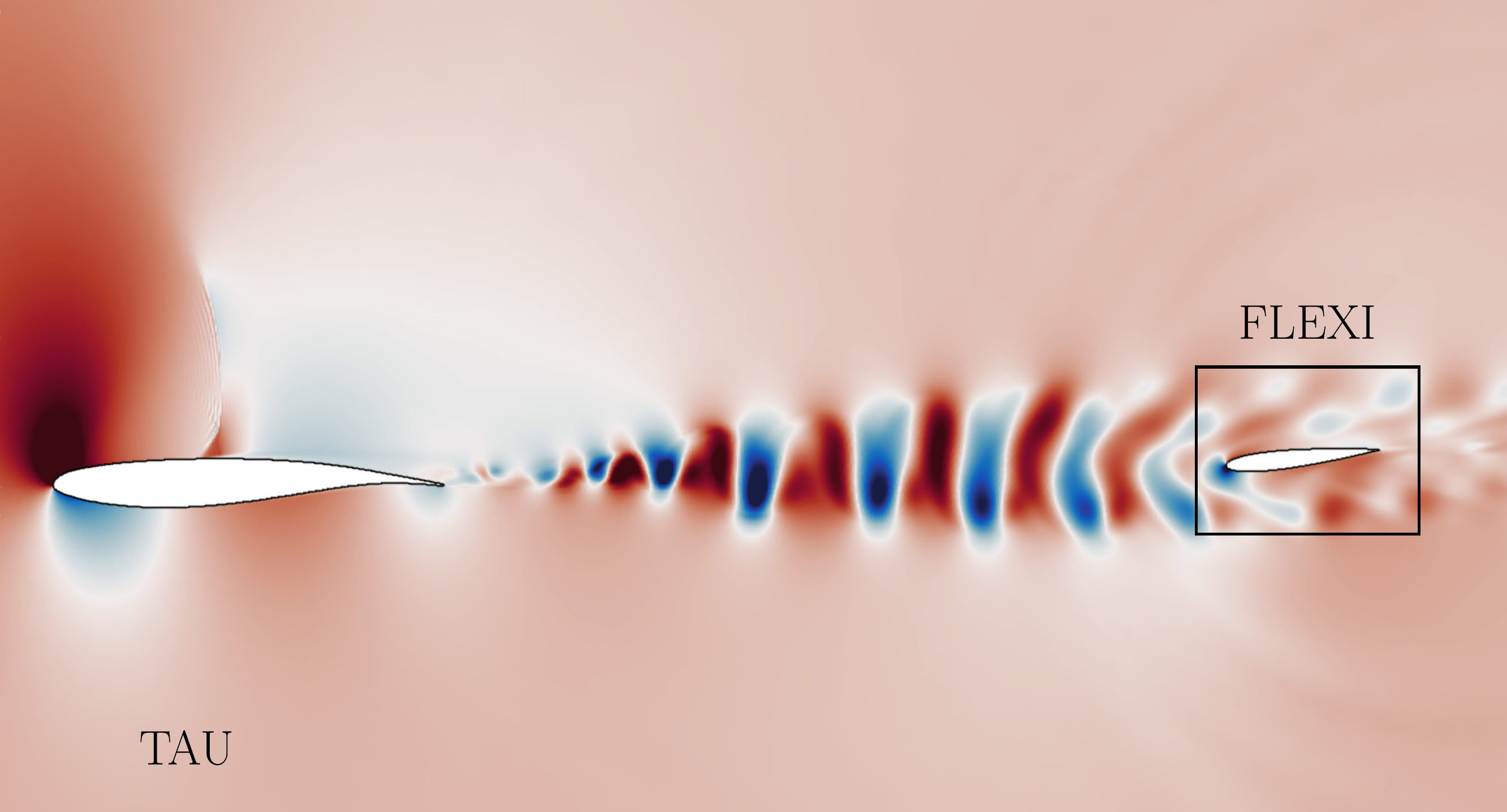}
	\caption{Tandem wing configuration with visualized wing wake interacting with the HTP simulated in TAU. Region of interest for detailed simulation in FLEXI is highlighted.}
	\label{fig:tandem_overview}
\end{figure}

In contrast to the spectral element code, the source data is typically point-wise data resulting from an experiment or finite volume code. The presented mapping algorithms are designed and implemented to be generally applicable, but are optimized to work with the DLR finite volume code TAU. In \fref{fig:tandem_overview} a typical application of TAU is visualized. TAU uses hybrid RANS/LES methods \eg for cases with separated flows, where attached boundary layers are treated in RANS mode and detached wake regions are resolved in LES mode. Thus the effect of the wing wake on the boundary layer of the HTP is hard to investigate within TAU. FLEXI on the other hand resolves the boundary layer and thus is capable of quantifying the influence of the wing wake. Thus, the region of interest that can be simulated within FLEXI is marked in \fref{fig:tandem_overview}. TAU will also be used to validate the results in \sref{results} later on.

\subsubsection{TAU}
The finite volume solver TAU is developed by the German Aerospace Center (DLR) and widely used among the aviation industry \cite{schwamborn2008}. It solves the Euler or RANS equations both on structured and unstructured grids. 
Several one- and two-equation models and Reynolds stress models are implemented for turbulence modeling. Additionally, LES or hybrid RANS/LES simulations can also be performed. 
Hexahedra, tetrahedra, triangular prisms and pyramids are supported elements for the cells of the primary grid. 
For the computation of the numerical fluxes at the cell boundaries, different upwind schemes and central approximations are available. 
Both explicit and implicit schemes can be chosen for the integration in time. The resulting linear system is solved with SGS or LUSGS schemes. 
For convergence acceleration, local time stepping, residual smoothing and multigrid methods are used. Parallelization is achieved by domain decomposition, with communication through the message passing interface (MPI).

\subsubsection{FLEXI}

FLEXI is a high-performance open-source CFD solver based on the Discontinuous Galerkin Spectral Element Method (DGSEM). It utilizes hexahedral tensor product elements with an arbitrary polynomial degree in each element. Since DG methods are hybrid schemes combining finite element and finite volume methods, we use a Roe Riemann solver with minimum dissipation for the fluxes between the elements \cite{Roe:1981}. Additionally, we represent the polynomial solution on a non-equispaced Legendre-Gauss or Legendre-Gauss-Lobatto point set. Since FLEXI acts as a framework there are multiple equation systems implemented. In this paper we mainly use the compressible Navier-Stokes equations. For validation the linear scalar advection system is used. The results in the application section are created using the compressible Navier-Stokes equations, which are implemented as skew-symmetric split form approximations to minimize aliasing instabilities \cite{Gassner:2018}. The boundary conditions generally are imposed weakly. This means, that we do not prescribe the state at the corresponding solution point in the element, but rather prescribe the numerical flux. FLEXI is parallelized using MPI and was successfully tested on up to $\mathcal{O}(10^5)$ cores \cite{Blind:2022}.

\subsubsection{Comparison of the Code Frameworks}

\begin{figure}
	\centering
	\includegraphics{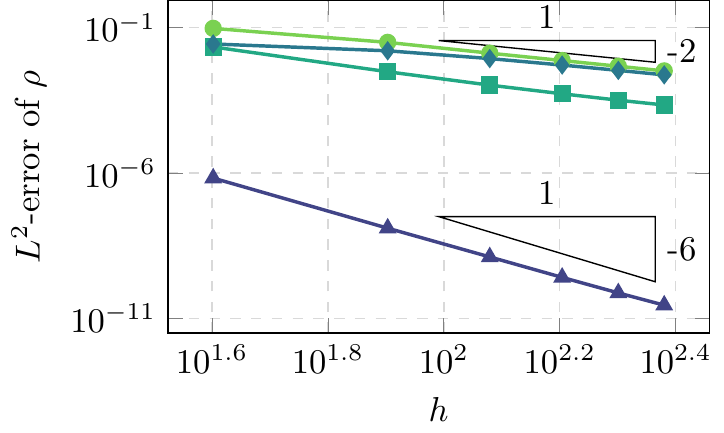}
	\includegraphics{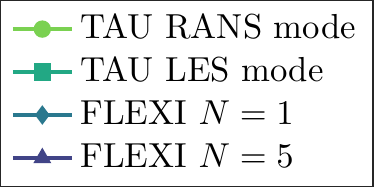}
	\caption{Comparison of the convergence behavior of TAU and FLEXI for different settings and meshes.}
	\label{fig:comparison_plot}
\end{figure}

Discontinuous Galerkin methods are commonly used high-order schemes. Finite volume methods in contrast are for unstructured meshes often limited to second order. It is well known that for the same number of degrees of freedom high-order methods can achieve lower error and need fewer solution points to resolve the same structures. This is known as numbers per wavelength $n_\text{PPW}$ criteria \cite{Gassner:2011,Flad:2020}. High-order methods can achieve $n_{PPW}\approx4$, while second finite volume method is often limited to $n_\text{PPW}\approx20$. This means that for resolving a structure of a given wavelength, high-order methods need 5 times fewer resolution points. However, this property is heavily dependent on the used polynomial degree $N$ as shown in \fref{fig:comparison_plot}. Thus, on the same mesh  the simulation with FLEXI $N=5$ not only has a faster decreasing error but also has the smaller error for a given amount of degrees of freedom. This becomes obvious since FLEXI $N=1$ or TAU on the $h=10^{2.4}$ grid correspond in terms of amount of degree of freedom with FLEXI $N=5$ at $h=10^{1.6}$. The Shu-vortex test case \cite{Spiegel:2015} is utilized to conduct this study. All simulations are run independently and thus without mapping.

The results denoted as ``TAU RANS mode'' are obtained with numerical settings commonly used for the RANS zones of hybrid RANS/LES simulations in TAU. A second order central flux approximation is used as Riemann solver stabilized by artificial dissipation, derived from the scheme after James, Schmidt and Turkel \cite{Jameson2017}. 
Applying a skew-symmetric scheme with matrix dissipation \cite{Swanson1992} already reduces the dissipation level compared to the TAU-default average of flux scheme with scalar dissipation. 
The simulations denoted with ``TAU LES mode'' additionally utilize a reconstruction of the convective fluxes using a linear gradient extrapolation at the cell faces, in a way to reduce the numerical dispersion of the skew-symmetric scheme \cite{loewe2016}. Moreover, the coefficient of artificial dissipation is lowered by a factor of 16. These settings are suitable for the LES zones of a hybrid simulation.
In the FLEXI runs a Roe Riemann solver is used \cite{Roe:1981}. The FLEXI simulation for $N=1$ shows a result similar to the TAU runs with the same order of convergence. However $N=1$ is typically not used in practical application, since the advantages of high-order schemes are not visible for such low polynomial degree. The runs with $N=5$ represents a more realistic scenario and show the advantage of high-order polynomials.

\subsection{Workflow}

\begin{figure}
	\centering
	\includegraphics[width=0.7\columnwidth]{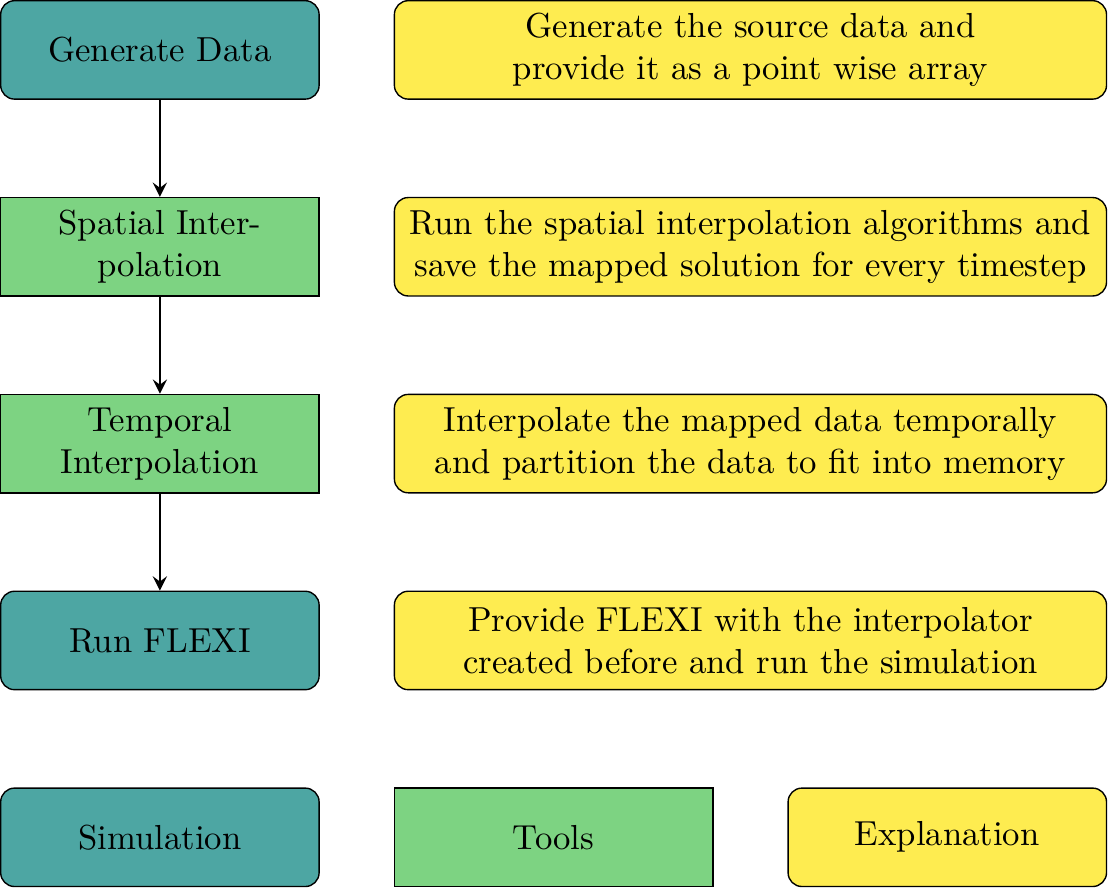}
	\caption{Workflow of imposing a time resolved boundary condition.}
	\label{fig:workflow}
\end{figure}

Before presenting the mapping routines, we first discuss the general workflow of how to run a simulation with time resolved input data. The general workflow is visualized in \fref{fig:workflow}.

First, the source data has to be provided. Generally this can be in the form of point-wise scattered data. Since in this paper we focus on the procedure for mapping TAU data to FLEXI we assume to get either volume snapshots or surface data from TAU. 

Second, we process the data by choosing an appropriate spatial mapping mechanism. Also we have to decide if we only want to map surface data for an instantaneous boundary condition, or if we also want to get the volume information to \eg generate a restart file for FLEXI. The tool creates an interpolated file for each input file. The results are saved in a HDF5\textsuperscript{\textregistered} format that uses a polynomial structure closely related to FLEXI. To ensure compatibility with FLEXI, the output polynomial degree is identical to the degree of the simulation we want to run afterwards. 

Higher-order representations are prone to aliasing and oscillations in general and the quality of the results heavily relies on the used point set and polynomial degree we perform the interpolation on. Since the output degree and point set is defined by the simulation we want to perform eventually, we can not use these parameters for mitigating errors. Thus, we super-sample the target point representation which helps avoiding errors due to oscillations resulting from the non-polynomials character of the source points. We then map the source data to the super-sampled target data points. We found that using $M\approx[1.5,2]\cdot(N+1)$ target points for super-sampling yields good agreement and mitigates oscillations significantly. This is in line with the literature values for overintegration of turbulent data, which is commonly used in the DG community for dealiasing \cite{Flad:2016}. After interpolating the source data to super-sampled target points, we project the solution to the original basis $N<M$ which removes the high modal information that is especially affected by aliasing.

Third, we interpolate the results from the second step temporally. The mapped volume or surface files created in step two get converted into a dataset containing the temporal interpolator for each solution point. The interpolator consists of the coefficients of the polynomial, which are dependent on the evaluation time. Additionally, we partition the data into a user-defined number of subsets to limit the amount of data of each interpolator and avoid memory overflow during FLEXI runtime.

Fourth, we provide FLEXI with the resulting file. FLEXI then evaluates the interpolant in each time step and sets the according boundary condition to the interpolated values.

\subsubsection{Some Remarks on Surface Data}

The mapping algorithms we present can be applied to volume as well as surface data. Depending on the provided data the spatial mapping algorithms will either use the volume solution to extract the target boundary or use the provided surface plane directly. 

TAU is able to write 2D data from a user-defined plane, onto which the flow variables are interpolated internally using algorithms of the chimera technique \cite{Burggraf:1999}. This interpolation will also of course introduce an error that leads to a mismatch between the volume solution of TAU and the 2D data on the plane. The resulting chimera plane can be read in separately. Hence, there can be made significant simplifications in term of area reduction which reduces the overall cost of the mapping algorithms. 

\subsection{Spatial Interpolation}

An important step to achieve the coupling of TAU with FLEXI is the spatial mapping. Since ultimately we want to create a instantaneous boundary condition we have to map surface data only. To keep the interpolation more general we implement a three-dimensional method to also allow volume interpolation and arbitrary oriented surface planes in the source domain. 

\begin{figure}[!htb]
	\centering
	\includegraphics[width=0.85\columnwidth]{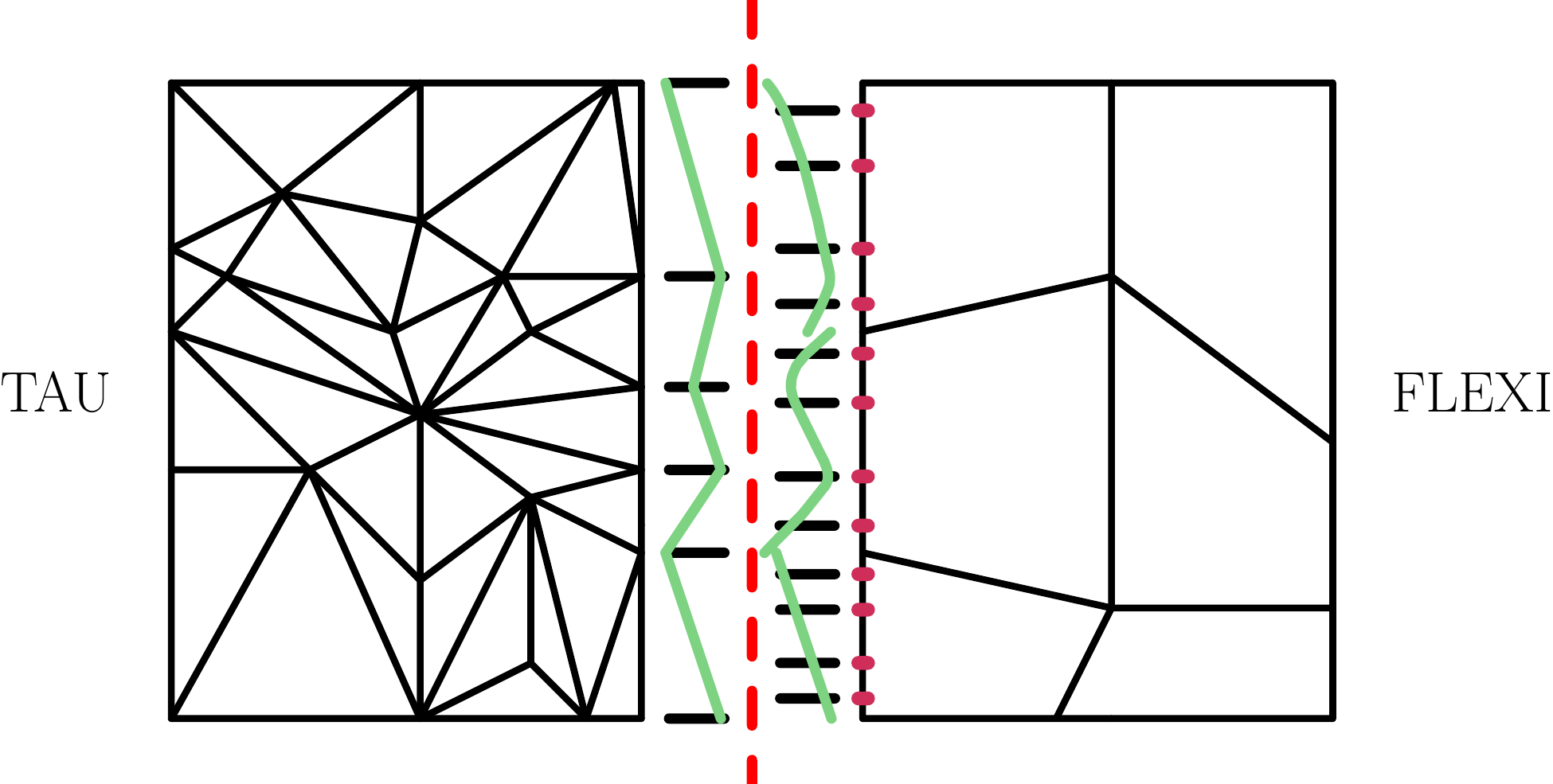}
	\caption{Visualization of the TAU-FLEXI interface including different point-sets. Light green curve shows an approximate solution after interpolating the TAU solution to the FLEXI point-set.}
	\label{fig:pointsets}
\end{figure}

A major challenge in creating a mapping between TAU and FLEXI are the differences in spatial discretization. Industrial finite volume codes rely often on tetrahedral meshes. Therefore, a direct interpolation is difficult, since mesh data structures for hexahedral only codes are dissimilar. In contrast to FLEXI, TAU stores its solution as low order point wise data. This results in arrays containing the coordinates $(x,y,z)$ and the solution $\vec{U}$ \cite{NetCDF:2022}. FLEXI however stores its solution as polynomial data for each element \cite{Krais:2021,Hindenlang:2014}. The difficulty now is to consistently map the point-wise TAU data to the polynomial coefficients needed for the solution polynomials in FLEXI. Since FLEXI only uses unstructured hexahedral elements and TAU is in contrast also based on tetrahedrals we can not use the TAU mesh information natively. An example of this incompatibility including a schematic of the interpolated solution is visualized in \fref{fig:pointsets}.

The interpolation algorithms thus have to work with scattered data. This means we have a cloud of points with a submerged target mesh. Therefore, interpolation from the TAU source data to the FLEXI target data has to be done using unstructured interpolation algorithms. There are several algorithms available that are suitable for such tasks. 

Scattered data interpolation generally can achieve good accuracy and performance but is highly dependent on the distribution of the source points \cite{Laughton:2020}. On the other hand implementing scattered data interpolation routines enable us to gain a more universal interface, since other codes and solution formats can be implemented easily.

Since the solution data of the source solver is provided beforehand, we do not have to map the data during run time, which saves a lot of computation time. In addition, the meshes used by both codes are known a priori (and remain constant during the computation). Still, good performance is crucial. Therefore, we implement the interface framework with MPI parallelization. 

\begin{figure}[!htb]
	\centering
	\begin{subfigure}[t]{0.49\textwidth}
		\centering
		\includegraphics[width=0.85\columnwidth]{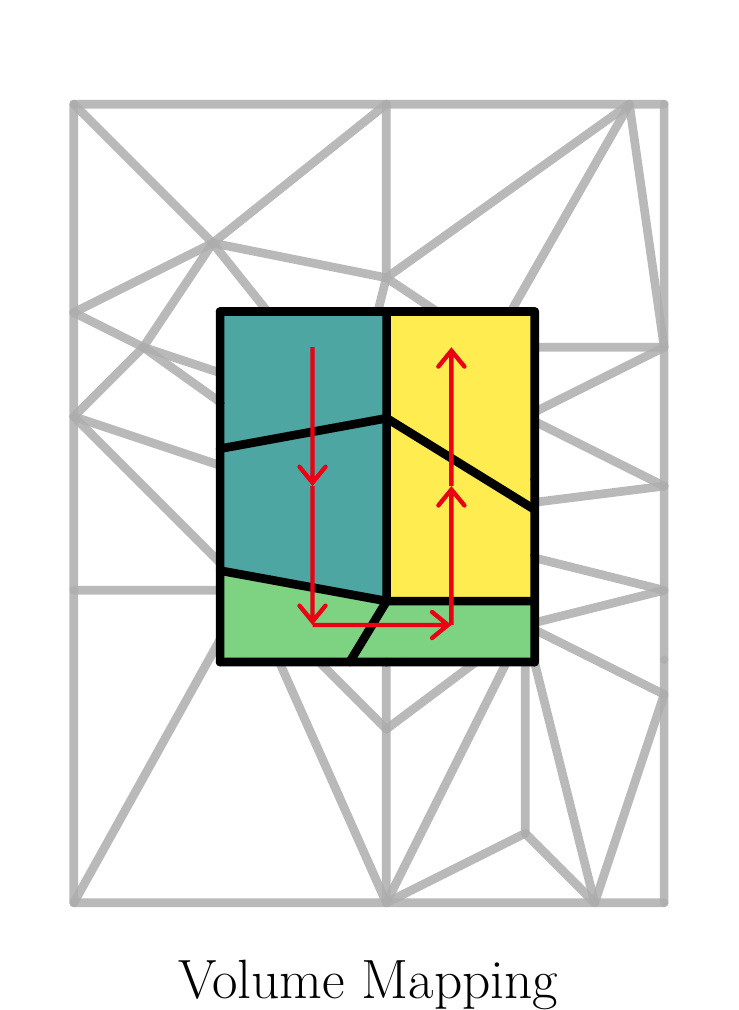}
		\caption{Target domain is fully submerged in the source data (gray). Domain decomposition does not have to take surface elements into account.}
		\label{fig:MPI_volume}
	\end{subfigure}
	\begin{subfigure}[t]{0.49\textwidth}
		\centering
		\includegraphics[width=0.85\columnwidth]{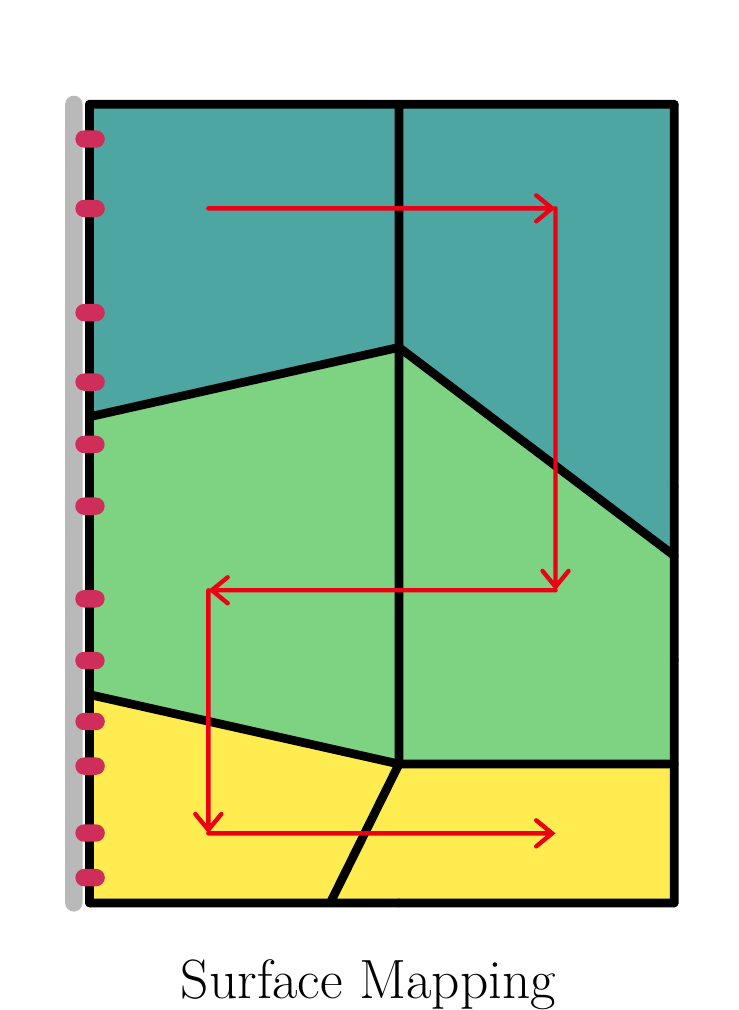}
		\caption{Boundary is aligned with surface source data (gray). MPI distribution has to be adapted in order for all three processors to contain mapped surface elements.}
		\label{fig:MPI_surface}
	\end{subfigure}
	\caption{Differences between MPI domain decomposition for volume and surface mapping on three MPI processors. Same colors correspond with the same MPI domain. Space filling curve is visualized in red.}
	\label{fig:MPI}
\end{figure}

This is done by reading in the mesh and distributing the elements equally between each processor using MPI. The elements are sorted along a Hilbert curve \cite{Hindenlang:2014} which minimizes the communication interfaces between the individual processors \cf red curve in \fref{fig:MPI_volume}. This approach however can only be done for the target FLEXI mesh, because we need the mesh information to distribute the data directly. For the source data we only read in data points. Hence, a simplification and distribution of the load between the processors is not possible in a first step. Thus, we read in the source data into shared memory \cite{Kopper:2022,Blind:2022}. Each processor then accesses the shared memory source data and simplifies it by only selecting the data which is necessary to spatially interpolate in its individual domain. With this method we can save a lot of computational time and decrease our memory footprint without sacrificing accuracy. 

If we use surface data as an input to the spatial interpolation algorithms we have to consider load balancing in more detail. The distribution of volume elements that has just been described, does not take surfaces into account. Thus, for the surface mapping we can not ensure that every part of the decomposed domain contains surface elements that have to be mapped (\cf \fref{fig:MPI_volume}). Therefore, especially for small domains with many processors, it is possible that not all processors are working on the task resulting in a inefficient mapping. Hence, we have to redistribute the load between the processors according to the number of surface elements (\cf \fref{fig:MPI_surface}). This can be done by assigning each surface element a high weight for domain decomposition. This weight is chosen by counting the number of boundary sides that have to be mapped per element and it generally reduces the computational load on MPI ranks that contain such a coupling interface. To do so we first have to read in the mesh file normally, then apply the surface weighting and finally reinitialize the mesh reading process \cite{Appel:2022}. With this approach we can gain performance improvements while sacrificing a few seconds in the initialization process due to the necessary reinitialization of the mesh. The overall cost of surface interpolation will be lower than using volume data. The difference between volume and surface distribution is visualized in \fref{fig:MPI}.

Additionally, we have to ensure to provide a buffer region around every individual MPI domain in order to establish the interpolation stencils for each point. The buffer area is estimated by taking the size of the largest element in the complete domain of the source points into account. Since the largest element is not known directly, we take the distances between the scattered points into account and use the largest distance for that matter.

\subsubsection{Nearest neighbor interpolation}
The nearest neighbor interpolation checks the source data coordinates and finds the closest point to the desired FLEXI target point by point-wise comparison. The values $U$ of the source data are then directly stored as a nodal coefficient in FLEXI. This type of interpolation yields a piecewise constant solution. Another requirement is an evenly distributed source mesh. If these requirements can not be met, there is risk of bad results. This does not automatically mean the the results are not physical, but rather that the resulting interpolation polynomial inside a DG cell is ill conditioned and can, due to the massive jumps, result in an unnaturally oscillating mapped solution. Another phenomenon one can observe is the possibility to get jumps on the element boundaries of the target mesh, if the boundary nodes are not included. On the other side this interpolation technique yields fast and good results if the source and target data are well aligned or if the meshes coincide at the interface.

\subsubsection{Inverse Distance Weighting}
A more general approach is available using inverse distance weighting \cite{Mueller:2004}. The target solution is calculated using a weighted average of the the source value
\begin{equation}
	u(\vec{x})_\text{target} = \dfrac{\sum_{i=i}^{N_\text{source}}\omega_i(\vec{x})u_{i,\text{source}} }{\sum_{i=i}^{N_\text{source}}\omega_i(\vec{x})}
\end{equation}
with $N_\text{source}$ denoting the number of source points in the whole domain. In contrast to the nearest neighbor approach we not only take one point into account, but all in the source area. The weights $\omega(\vec{x})$ are depending on the distance between the source points and the target solution point
\begin{equation}
	\omega_i = \dfrac{1}{\left(\norm{\vec{x}-\vec{x}_i}\right)^p}
\end{equation} 
and a weighting exponent $p$. For $p\Rightarrow\infty$ the inverse distance weighting approach resembles the nearest neighbor method. A modification to the general inverse distance weighting was introduced by Shepard, who proposed to only take the source points into account that are within a predefined radius $R$ around the target point \cite{Shepard:1968}. This reads as
\begin{equation}
	\omega_i = \left(\dfrac{\max(0,R-\norm{\vec{x}-\vec{x}_i})}{R\left\|\vec{x}-\vec{x}_i\right\|_2}\right)^2
\end{equation} 
with $R$ denoting a predefined search radius.

\subsubsection{Radial Basis Functions}
A third option to consider for unstructured interpolation are radial basis functions $\varphi$ \cite{Hardy:1971,Hardy:1990}. These methods allow for high-order accurate interpolants $s$ of unstructured data. The interpolant consists of the weighted sum of radial basis functions. In contrast to the other methods introduced earlier we have to solve a linear equation system to invert the Vandermonde and to determine the weights $\omega$ satisfying
\begin{equation}
	s(\vec{x}) = \sum_{i=1}^{N_\text{source}}\omega_i\varphi(\norm{\vec{x}-\vec{x}_i})
\end{equation}
and therefore 
\begin{equation}
	u_{j,\text{source}} = \sum_{i=1}^{N_\text{source}}\omega_i\varphi(r_{ji})
\end{equation}
with $r_{ki}=\norm{\vec{x}_k-\vec{x}_i}$. We rewrite the interpolation condition in matrix notation
\begin{equation}
	\begin{bmatrix}
		\varphi(r_{11}) & \varphi(r_{12}) & \cdots & \varphi(r_{1N}) \\
		\varphi(r_{21}) & \varphi(r_{22}) & \cdots & \varphi(r_{2N}) \\
		\vdots    & \vdots     & \ddots & \vdots\\
		\varphi(r_{N1}) & \varphi(r_{N2}) & \cdots & \varphi(r_{NN})
	\end{bmatrix}
	\begin{bmatrix}
		\omega_1 \\
		\omega_2 \\
		\vdots \\
		\omega_N
	\end{bmatrix}
	=
	\begin{bmatrix}
		u_{1,\text{source}} \\
		u_{2,\text{source}} \\
		\vdots \\
		u_{N,\text{source}}
	\end{bmatrix}.
\end{equation}
This can be rewritten in matrix form as $\mathbf{\Phi}_{ij}\vec{\omega}_i=\vec{u}_{j,\text{source}}$ using index notation. Since we have to invert the matrix $\mathbf{\Phi}$ for interpolation, the radial basis approach is the most expensive of the introduced methods. 

We evaluate the interpolant 
\begin{equation}
	u(\vec{x}) \approx \sum_{i=1}^N \omega_i \varphi(\norm{\vec{x} - \vec{x}_i})
\end{equation}
and get the value at an arbitrary point in the computational domain.

Typical radial basis functions for interpolation are multiquadratic $\varphi(r) = \sqrt{1+(\varepsilon r)^2}$, inverse multiquadratic $\varphi(r) = \frac{1}{\sqrt{1+(\varepsilon r)^2}}$, Gaussian $\varphi(r) = e^{-(\varepsilon r)^2}$ and thin plate spline $\varphi(r) = r^2\ln(r)$ functions with $r=\norm{\vec{x}_j-\vec{x}_i}$. The parameter $\varepsilon$ defines the shape of the function and is used for scaling. The multiquadratic and the thin plate spline have shown to be the most reliable radial basis functions for this use case. Since the thin plate spline does not require any additional user parameter $\varepsilon$ we use this function for all further investigations.

During implementation of the algorithms above some observations were made. First, none of the scattered interpolation method is designed in a way to be conservative. Thus, we interpolate the primitive variables and, for consistency reasons, convert to conservative variables after mapping.

\subsubsection{Comparison of the Spatial Interpolation Methods}

Before assessing the performance of the spatial interpolation routines in context of the mapping routines, we investigate the convergence behavior in an isolated test case. Thus, we calculate the $L_2$-error of a simple one-dimensional interpolation of a sine function $f(x)=\sin(2\pi x)$.

\begin{figure}
	\centering
	\includegraphics{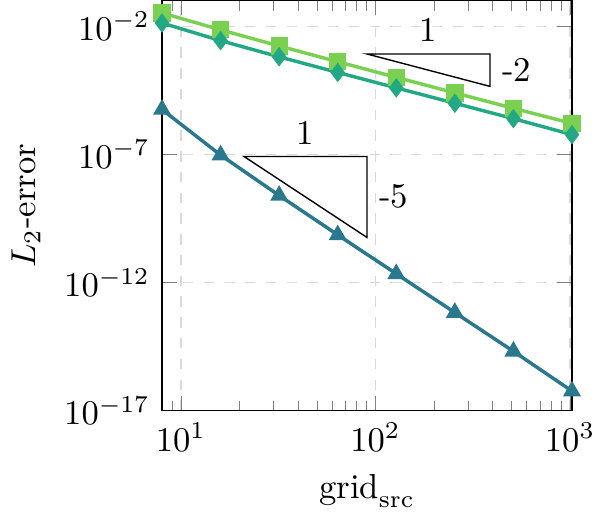}
	\includegraphics{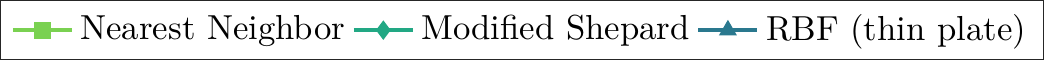}
	\caption{Convergence of the $L_2$-error of an interpolated one-dimensional sine function for different interpolation methods.}
	\label{fig:interpolation_convtest}
\end{figure}

We plot the error in \fref{fig:interpolation_convtest} over the sampling resolution of the source data. We can see that radial basis function interpolation clearly yields the best results with lower errors and a better convergence rate $EOC=5$ than nearest neighbor and inverse distance weighting interpolation with $EOC=2$. We assess the accuracy and the differences between the spatial interpolation schemes in more detail in \sref{validation}.

\subsection{Temporal Interpolation}

In addition to the spatial interpolation we also have to interpolate temporally in order to account for the different time stepping schemes in the source and target codes. FLEXI \eg uses an explicit low storage Runge-Kutta method to advance the equation systems in time. TAU on the other hand uses an implicit time discretization to accomplish that. However, in addition to the different time stepping schemes, the time step and output rate of the simulation data can change between different simulations. For using the data as an instantaneous boundary condition we have to ensure that we can provide the target solver with the correct inflow data at an arbitrary point of time. Thus, it is crucial to interpolate the results of the spatial interpolation in time to get a continuous temporal interpolator.

In contrast to the volume and surface mapping the temporal interpolation consists of purely one-dimensional problems. For one-dimensional data there are vast numbers of different interpolation techniques. In this work we use polynomial interpolation in combination with a Lagrange basis and spline interpolation. 

We use the Lagrange interpolation basis since coefficient and solution values coincide \cite{Kopriva:2009}. Also the evaluation can be easily done with the tools already built into FLEXI, since the solution in each element consists of the tensorproduct of three one-dimensional nodal Lagrange functions. 

Furthermore two different variants of spline interpolation are implemented. A common open spline as well as the Akima spline \cite{Akima:1970}. In contrast to a typical spline an Akima spline does not take the second derivative into account. This leads to a more evenly distributed solution and fewer overshoots compared to the open spline. The problem of overshoots can also be found in polynomial interpolation of degree $p\geq 2$. This becomes especially important if an implicit source method is paired with an explicit target solver. In this case the temporal interpolation has to come up for a huge number of time steps since the time step in an explicit scheme is typically much smaller than implicit time steps. Thus, overshoots can play a significant role for the overall mapping quality. If the time steps of source and target method are similar, the effect of the temporal interpolation becomes smaller. However, it should be noted that even in this case, overshoots can be generated. Generally, for these reasons it is recommended to either use linear interpolation or the Akima interpolation for the most reliable results. We will show this in \sref{validation}.

Hence, the resulting quality of the interpolation depends on multiple factors. First, on the chosen interpolation method. Second, on the sample rate of the provided state or boundary source files. Thus, a general prediction of the error resulting from temporal interpolation is difficult.

The interpolation is done in a separate tool and is not only limited to surface data, but can also be done with restart files of any FLEXI simulation. The result is processed and saved in an HDF5\textsuperscript{\textregistered} file which includes the coefficients for every polynomial at every temporal sample point.

The resulting files of the temporal interpolation routine can either be directly used in FLEXI for evaluation of the interpolant or even be used to generate a restart file to continue simulation at an arbitrary point of time.

The temporal interpolator generated contains the resulting polynomial/spline at each degree of freedom. Thus, the overall size of the interpolator array has more dimension (polynomial coefficients and time)  than the solution array. With increasing dimensionality the memory requirements of the array also increase. Depending on the amount of source data available it might be necessary to partition the resulting temporal interpolant in order to avoid memory overflow during simulation of the target domain. Therefore, a maximal size for the interpolant array has to be provided by the user. The interpolation algorithm will then partition the data into equally sized datasets, each containing a period of time which results from the user parameter. FLEXI then only reads in the dataset that contains the temporal information of the current FLEXI time step. Thus, during runtime of FLEXI the saved interpolant is only evaluated, allowing for obtaining an interpolated solution at an arbitrary point of time.
 \section{Validation of the Interface}\label{validation}

In this section we start validating the mapping algorithms. We chose a gradual approach and start by showing a proof of concept, followed by the temporal algorithms and in the end assess the convergence behavior of the spatial interpolation algorithms of the interface.

We start to evaluate the algorithms by applying them to very simple test cases. Thus, we chose the linear scalar advection equation
\begin{equation}
	u_t+\nabla\cdot(au) = 0 \quad\text{ with }\quad a \in \mathbb{R}
	\label{eqn:linscalaradvection}
\end{equation}
 due to its simplicity and a priori known exact solution for given initial conditions. The transport velocity is set to $a=1$. We vary the initial conditions between the tested scenarios and describe them in the corresponding sections in more detail. The source domain $\Omega\in[-1,1]^3$ and the target domain $\Omega\in[1,3]\times[-1,1]^2$ are designed to have a shared interface at $x=1$. The source data as well as target data for these test cases are fully generated using FLEXI. Triple-periodic boundary conditions are used for the source mesh and the target mesh is designed to have periodic boundary conditions in $y$ and $z$. In $x$-direction we have the instantaneous interface condition at $x=1$ and a outflow at $x=3$.

\subsection{Proof of Concept}

We start the validation of the interface by applying it to a very basic sine test case with 
\begin{equation}
	u(x,t) = sin(\pi(x-at)).
\end{equation}
The initial conditions are set to $u_0(x,t=0)$. The exact solution $u(x,t)$ is purely $x$ dependent and thus the values at the interface plane $u(x=1,t)$ do not vary in $y$ and $z$. The target domain is initialized with a constant solution $u_0 = \text{const.} = 0$. 

\begin{figure}
	\centering
	\includegraphics{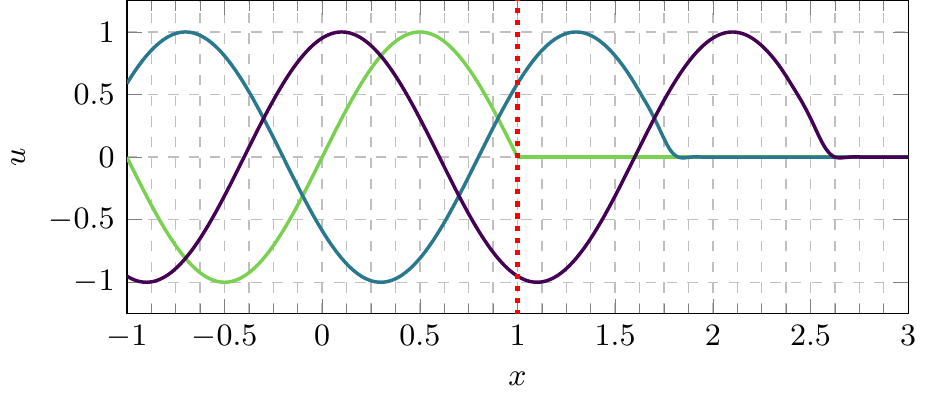}
	\includegraphics{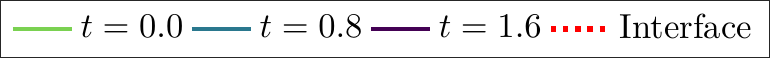}
	\caption{Overview of the spatial mapping process for a traveling sine wave.}
	\label{fig:spatial_sine}
\end{figure}

For this first test we match the surface elements at the interface and thus can use nearest neighbor interpolation without sacrificing accuracy (source and target points coincide). In this case the nearest neighbor algorithm will just copy the data from the source to the target domain. Source and target mesh are only offset in $x$-direction by the length of the domain. In \fref{fig:spatial_sine} the initial condition is depicted in light green. One can also see the solution of \eqref{eqn:linscalaradvection} after $t=0.8$ and $t=1.6$. The vertical gray dashed grid lines depict the mesh of the simulation grid and the red dotted line visualizes the interface between source and target domain. The graphs are extracted from the center line in $x$-direction of the equispaced Cartesian cubes, which each have a resolution of $16\times16\times16$ using $N=4$ polynomials in each element. Legendre-Gauss-Lobatto points are used for the source and target simulation. Additionally, we avoid temporal interpolation by sampling the interface data at every physical time step $dt$.

In \fref{fig:spatial_sine} we can see that the general workflow presented performs as expected and the information gets propagated over the interface with $a=1$. Since we do not interpolate the data in any way in this test case we expect the overall errors between source sine and target sine wave to be comparable. In the source domain we have an $L_2$-error of $\num{9.6506E-07}$ after $t=2$. The $L_2$-error in the target domain after $t=2$ is evaluated in the same way and is $\num{9.6859E-07}$. This successfully proves that the workflow is capable of mapping the data without any information loss. We stress that the full framework is working as if we were coupling between two heterogeneous solvers, with the exception that the source and target points coincide. 

Note that we used continuous initial conditions between source and target domain. We found that one should avoid having jumps or large discrepancies of the initial conditions between the source and target domain due to nonphysical disturbances created at the inlet of the target domain which are further propagated. This however, is not due to the interface mapping algorithms but rather due to the nature of the high-order scheme. In practical applications, especially for transient simulations, this does not pose a problem since all structures starting from free-stream, will be mapped into the target domain.

\subsection{Assessing the Temporal Interpolation and Sampling}

\begin{figure}[!htb]
	\centering
	\begin{subfigure}{\textwidth}
		\centering
		\begin{center}
			\includegraphics{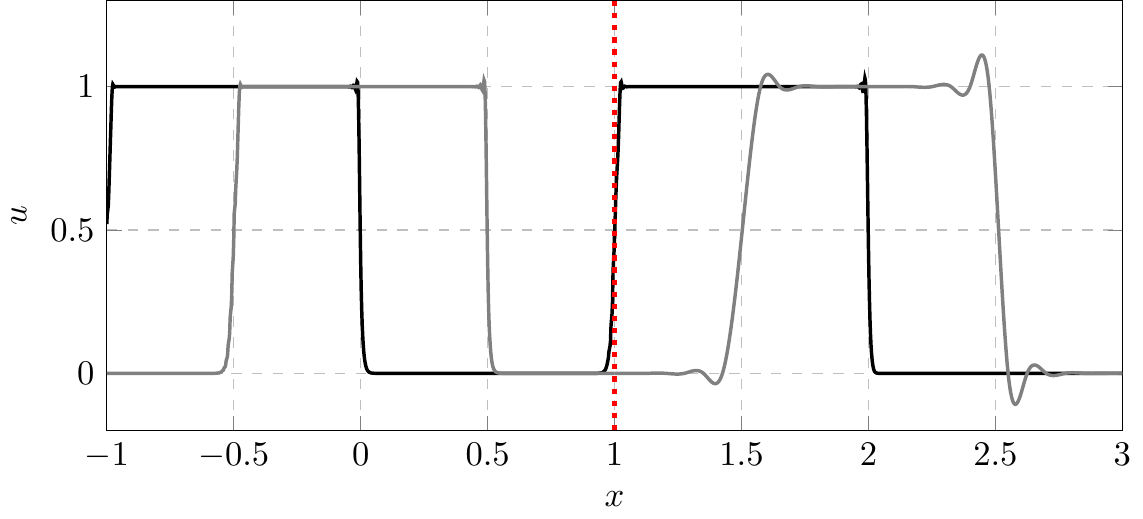}
			\includegraphics{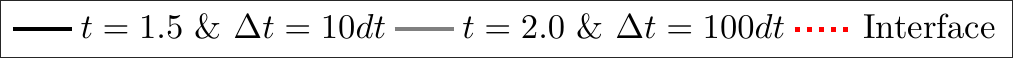}
		\end{center}
		\caption{Overview over the simulation domain of the jump test case using spline interpolation. The jumps are shown in more detail in \fref{fig:sampling_detail}.}
		\label{fig:sampling_overview}
	\end{subfigure}
	\hfill
	\begin{subfigure}{\textwidth}
		\begin{center}
			\begin{minipage}{0.32\textwidth}
				\centering
				\includegraphics{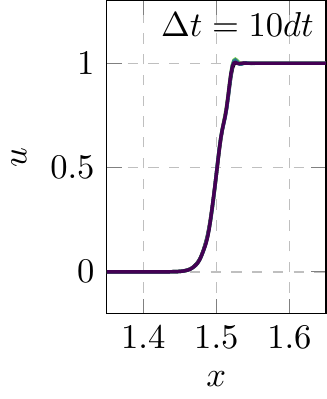}
			\end{minipage}
			\begin{minipage}{0.32\textwidth}
				\centering
				\includegraphics{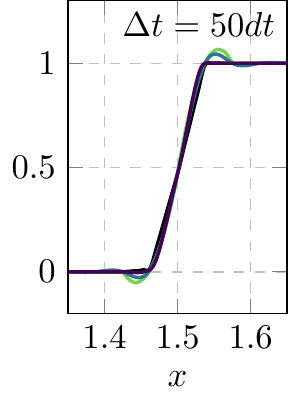}
			\end{minipage}
			\begin{minipage}{0.32\textwidth}
				\centering
				\includegraphics{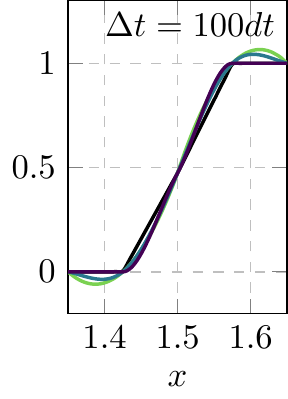}
			\end{minipage}
			\includegraphics{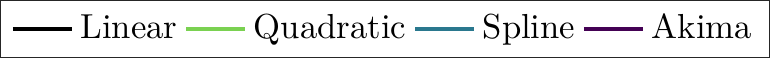}
		\end{center}
		\caption{Detailed view of the jumps containing the interpolation techniques visualized in \ref{fig:sampling_overview}.}
		\label{fig:sampling_detail}
	\end{subfigure}
	\caption{Overview and detailed plots of the jump test case for different $\Delta t$ and temporal interpolation algorithms.}
	\label{fig:sampling_mapping}
\end{figure}

Next, we evaluate the effect of temporal interpolation/sampling on the interface mapping process. Thus, we investigate the effects of different temporal interpolation schemes and sampling rates on the incoming solution, which we map via the instantaneous boundary condition. For this test case we chose different exact solution and initial conditions for the linear advection equation \ref{eqn:linscalaradvection}. To evaluate the effect of the sampling we chose a initial condition that includes a discontinuity in order to visualize the information loss. Thus, we use 
\begin{equation}
	u(x,t) = 
	\begin{cases}
		1. \quad\text{if}\quad -0.5<x-at<0.5 \\
		0. \quad\text{else}
	\end{cases}.
\end{equation}
We use Legendre-Gauss-Lobatto nodes with $N=4$ on a $256\times1\times1$ source and target domain. The surface elements at the interface are again matched. Thus, we can use nearest neighbor interpolation to interpolate the surface data in space, without sacrificing accuracy (copy values from source to target). \fref{fig:sampling_overview} depicts the simulation at two discrete points in time evaluated with different $\Delta t$-interpolants. The interface is located at $x=1$ and the discontinuities travel into the target domain on the right side of the red dotted interface with $a=1$. 

Already in the overview graph in \fref{fig:sampling_overview} we can see substantial differences between the two interpolated jumps. For this test we chose in total three sampling rates. A fine sampling rate at $\Delta t\approx10dt$ which is approximately ten times the explicit FLEXI time step $dt$ and two coarser sampling rates at $\Delta t\approx50dt$ and $\Delta t\approx100dt$. In \fref{fig:sampling_overview} only the finest and the coarsest sampling rate are visualized. \fref{fig:sampling_detail} shows the jumps of all three sampling rates at the same evaluation time $t$ in more detail. Since the $x$-axis is scaled identically one can see that the influence of the temporal mapping on the target FLEXI simulation is very high. There are two main effects visible:
\begin{enumerate}
	\item The temporal distance between two samples effects the slope of the jump and
	\item the temporal interpolation method has an effect on the quality of the jump representation.
\end{enumerate} 
The first observation has to only result from the temporal interpolation since the slope of the shock has been steeper in the source domain. Additionally we see that lowering $\Delta t$ increases the slope again. Thus, $\Delta t$ has to be chosen in a way that the steepest gradient in data can be represented sufficiently. This however is very much problem depending and requires knowledge of the data.  In \sref{results} we assess an approach on how to determine this in the context of turbulent eddies.

For the second point a more general statement can be made, since this observation is nearly independent of $\Delta t$ and only becomes more prominent if $\Delta t$ becomes sufficiently large. Higher order polynomial approximations tend to oscillate, especially for equispaced point distribution which is the case for temporal sampling. Therefore, in \fref{fig:sampling_mapping} polynomial interpolation is only shown up to second degree. Also the well known Spline interpolation tends to oscillations for high $\Delta t$. Most favorable thus is linear or Akima interpolation, which represent the vertical jump best and recover the steepest gradients. Higher order polynomial and spline interpolation in this case fail mainly because the physical time steps $dt$ at which we sample are roughly equispaced. Thus, we see the so-called Runge's phenomenon for interpolation using an equispaced point set in time. This is a crucial point since the TAU output frequency is only determined by its implicit timestep. The target solver FLEXI thus has to recover the data in every explicit time step. However, having a fixed source sampling rate and decreasing the target time step, will not further increase the error since the interpolant is only determined by the sampling rate of the source data and only is evaluated during FLEXI runtime.

Since Akima interpolation yields slightly smoother results in combination with steeper gradients, we use Akima interpolation for all following test cases.

\subsection{Convergence of the Spatial Mapping}

Another important aspect one has to consider is the convergence behavior of the mapping process. To measure the effect of the interpolation routines we decided to calculate the error norm of the whole mapping and simulation process. Thus, the error includes spatial and temporal interpolation error as well as the error associated with imposing the instantaneous boundary condition in FLEXI (\eg discretization error). 

To run the convergence test, we modify the initial conditions from the one-dimensional sine in \fref{fig:spatial_sine}. We add a $y$ and $z$ dependency to the exact solution $u$ in order to have varying $u$ values on the interface plane. Thus, we get
\begin{equation}
	u(x,y,z,t) = sin(\pi(x-at))+sin(\pi y)+sin(\pi z).
\end{equation}

For the sine wave and the linear transport we have seen earlier that we can recover the exact solution on the target domain and that the information is propagated correctly via the instantaneous boundary condition if there is no spatial and temporal interpolation involved (just copy the values from source to target). Thus, we want to investigate the effect of different non-matching interfaces (point sets and resolution) on the error of the simulation and therefore have to combine spatial and temporal interpolation techniques for the first time. We again use Legendre-Gauss-Lobatto points with $N=5$ in the source and target domain. Additionally, we use super-sampling with $N=8$. For the linear transport this is not necessary, since in contrast to the Navier-Stokes equations we do not see aliasing here. However, we want to assess the convergence as close to the later application as possible and additionally avoid matching all the degrees of freedom in any case ($N_\text{src}=5\neq8=N_\text{tar,super}$). 

\begin{figure}
	\centering
	\begin{subfigure}{\textwidth}
		\centering
		\begin{minipage}{0.49\textwidth}
			\centering
			\includegraphics{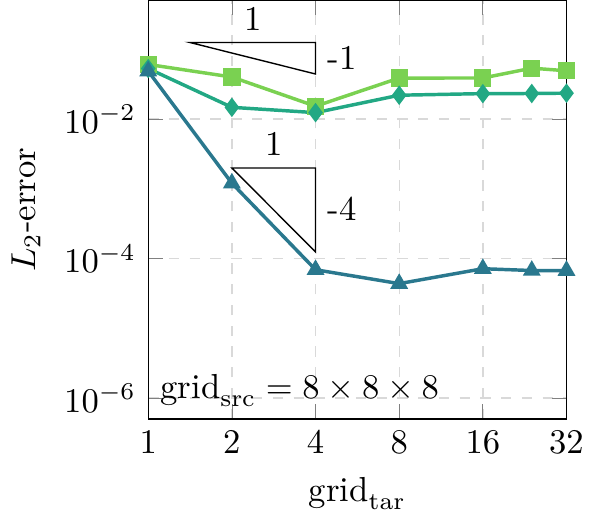}
		\end{minipage}
		\begin{minipage}{0.49\textwidth}
			\centering
			\includegraphics{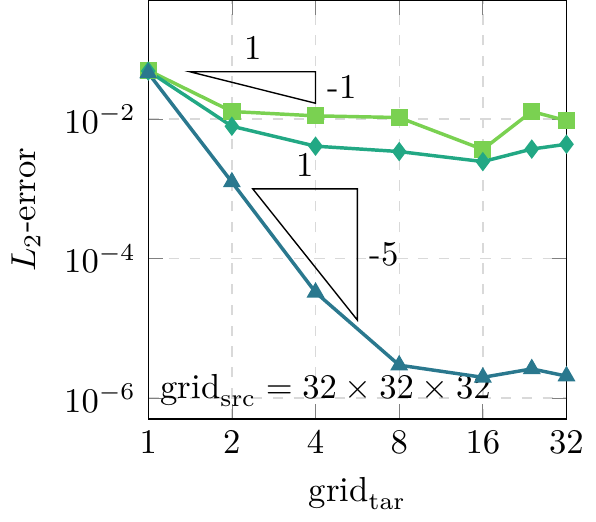}
		\end{minipage}
		\caption{Convergence behavior for the linear scalar advection equation system for two different source meshes.}
		\label{fig:convergence_source}
	\end{subfigure}
	\hfill
	\begin{subfigure}{\textwidth}
		\begin{minipage}{0.49\textwidth}
			\centering
			\includegraphics{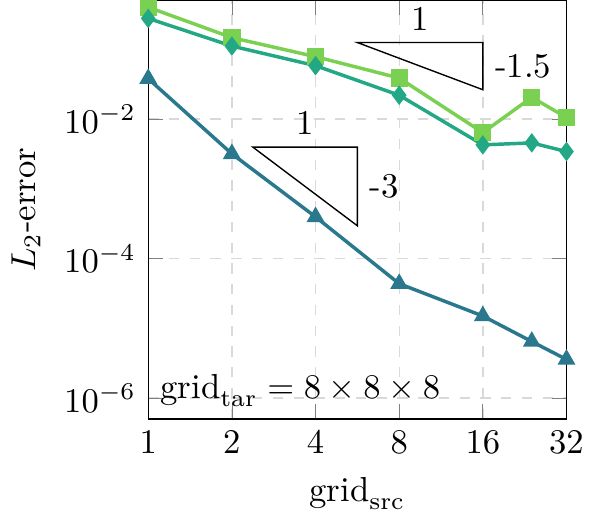}
		\end{minipage}
		\begin{minipage}{0.49\textwidth}
			\centering
			\includegraphics{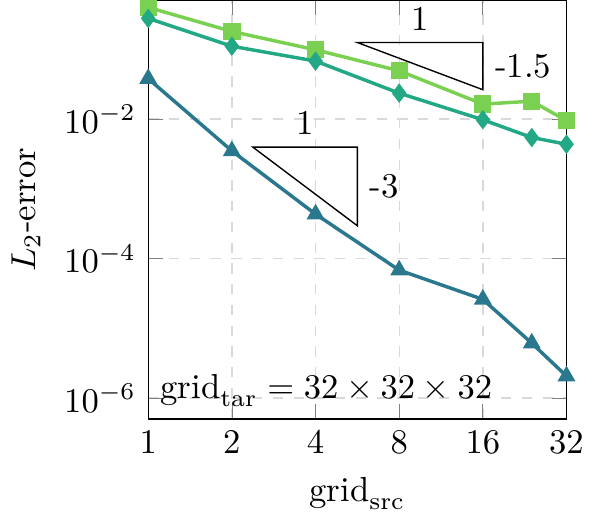}
		\end{minipage}
		\caption{Convergence behavior for the linear scalar advection equation system for two different target meshes.}
		\label{fig:convergence_target}
	\end{subfigure}
	\includegraphics{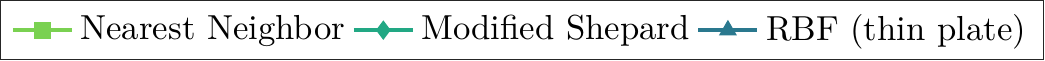}
	\caption{Comparison of the convergence behavior of the entire mapping procedure including spatial, temporal mapping and the instantaneous boundary condition.}
	\label{fig:convergence_interface}
\end{figure}

In \fref{fig:convergence_interface} we see two different testing scenarios. The first in \fref{fig:convergence_source} shows the $L_2$-error for increasing target resolution and a fixed source mesh. The second scenario in \fref{fig:convergence_target} depicts the error for an increasing source resolution and a fixed target mesh. With ``grid'' we mean the number of elements in each spatial direction of the Cartesian cube. The sampling timestep is defined by the physical timestep $dt$ of the source data. Thus, for \eg the source grid ``1'' we extract the interface data at every physical time step and use Akima interpolation to interpolate it to the physical time step of the ``32'' target grid that is 32 times smaller.

In \fref{fig:convergence_source} we assess the effect of varying target mesh resolutions on the overall error. The resolution of the source mesh is fixed at $8\times8\times8$ and $32\times32\times32$ respectively. We expect the overall error to converge, since the error can not be mitigated any further if it is dominated by the source data. Thus, we can see the influence of the source data on the target domain. For the $8\times8\times8$ source mesh we can observe this behavior very well. Starting at $\text{grid}_\text{tar}\approx4$ we see that for all interpolation algorithms there is no further improvement. For the finer source mesh we can observe a similar behavior, however the overall error is lower and the error is converged later. Due to the error introduced with the spatial interpolation we can not see a declining error until the source mesh resolution.

In \fref{fig:convergence_target} we investigate the effect of a varying source mesh on a fixed target mesh. This can be interpreted as increased input quality for the mapping for a given target domain. In this test case we again assessed the influence for two fixed target resolutions at $8\times8\times8$ and $32\times32\times32$. Nearest neighbor, Shepard as well as RBF interpolation show declining errors for increasing source grid resolution. This time nearly linear decaying errors can be seen up to the resolution of the target mesh. However, especially for the $\text{grid}_\text{tar}=8\times8\times8$ case, we can see that RBF interpolation is capable of recovering information from source grids with finer resolution than the target mesh. Shepard and nearest neighbor show clearly weaker performance here and have changing slopes of the error in this source grid regime.

Overall we can rank the performance of the three tested spatial interpolation techniques. Nearest neighbor interpolation shows as expected the weakest performance with an experimental order of convergence of $EOC\approx1.2$ in \fref{fig:convergence_target}. Shepard interpolation shows overall lower errors at roughly the same order of convergence $EOC\approx1.5$. However, Shepard interpolation is capable of retaining the error even for source resolutions higher than the target resolution in \fref{fig:convergence_target} where nearest neighbor interpolation show inconsistent results. Finally, radial basis function interpolation clearly yields the best results with an order of convergence of $EOC\approx2.8$. Thus, using RBF interpolation is recommended. 
Overall the results in \fref{fig:convergence_target} underline the observations made in \fref{fig:interpolation_convtest}. However, the convergence rates in \fref{fig:convergence_target} are lower for all interpolation methods. The qualitative observations however are identical and the losses in $EOC$ are equivalent for all interpolation techniques. One should note, that the test case in \fref{fig:convergence_target} has an increased complexity, since it is two-dimensional and we evaluate the error over the whole mapping process compared to an isolated one-dimensioal interpolation test in \fref{fig:interpolation_convtest}.

For large source datasets one should keep in mind, that solving the equation system necessary to the get the interpolation coefficients for the radial basis functions gets very expensive and RBF interpolation even in this simple test case was noticeably (approx.~up to an order of magnitude) slower than nearest neighbor and Shepard interpolation.
 \section{Results: Cylinder Flow}\label{results}

In this section we investigate the flow around a cylinder at a Reynolds number of $Re_c=3900$ \cite{Kravchenko:2000,Parnaudeau:2008}. The diameter of the cylinder is defined as $c$ and is used as the characteristic length in this investigation. The domain has a spanwise extension of $c$. For the first time we now map actual TAU surface data into a FLEXI domain.

\begin{figure}[!htb]
	\centering
	\includegraphics{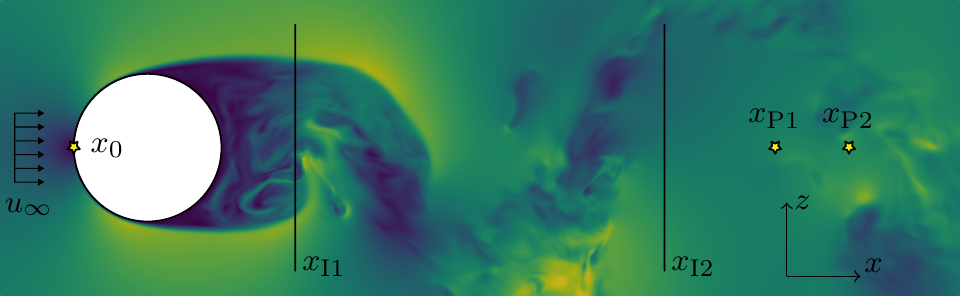}
	\caption{Cylinder at $Re_c=3900$ test case definition containing interface planes and probe points for evaluation.}
	\label{fig:cylinder_testcase}
\end{figure}

In \fref{fig:cylinder_testcase} the simulation setup is depicted. Note that the size of the interface planes in \fref{fig:cylinder_testcase} at $x_I$ does not match the size in the actual simulation. In the setup the interface planes are designed in a way that all the turbulent wake structures are captured by the planes and all vortical structures of the wake are fully contained in the interface planes.

The most important aspects for assessing the performance of the interface is to define the interface locations and to define record points (also known as probe points). We decide to place two interface planes at position $x_I$ in the wake as well as two record points $x_P$. 

We use the same number of degrees of freedom in the TAU source and the appended FLEXI target mesh. Thus, the target mesh resolution including the interface has to be divided by a factor of eight in order to accommodate for the higher polynomial degree of FLEXI $N=7$. With this approach we minimize the resulting errors (\cf \fref{fig:convergence_source}). We will show later that this resolution is sufficient to map all physical structures occurring in this test case. The target mesh in this case is a simple box that has the same $y$ and $z$ dimensions as the interface and a sufficiently long $x$ extension for the turbulent wake to develop and travel.

\subsection{Simulation Setup}

Next, we discuss the simulation setup of the cylinder test case. We describe the setup for FLEXI as well as TAU.

The main reason we chose the cylinder flow as the main test case is the fact, that we can afford to run the whole domain in FLEXI and in TAU. Thus, we not only can compare the mapped results against the TAU solution but also against the reference DNS created with FLEXI. Additionally, the cylinder is a well known geometry in the fluid mechanics community and has been investigated in detail before.

We use the same base mesh setup for the TAU and FLEXI DNS reference simulation. The only difference is again that the FLEXI case is coarser by a factor of eight to consider the high-order polynomials that are used in each element. Thus, if FLEXI is run with $N=7$ we have the same number of degrees of freedom as TAU. We also investigate the solution quality of lower order polynomials later. For the simulation in FLEXI we use $N\in[3,5,7]$ polynomials.

\subsection{Sensitivity on Resolution}

First, we assess the sensitivity of the test case on the resolution. We conduct this study in FLEXI since we are interested if the chosen resolution is sufficient for a DNS. This test case was specifically chosen since it allows to conduct a fully resolved simulation in FLEXI and in TAU. For typical applications of the inflow condition this will not be possible.

\begin{figure}[!htb]
	\centering
	\includegraphics[width=\columnwidth]{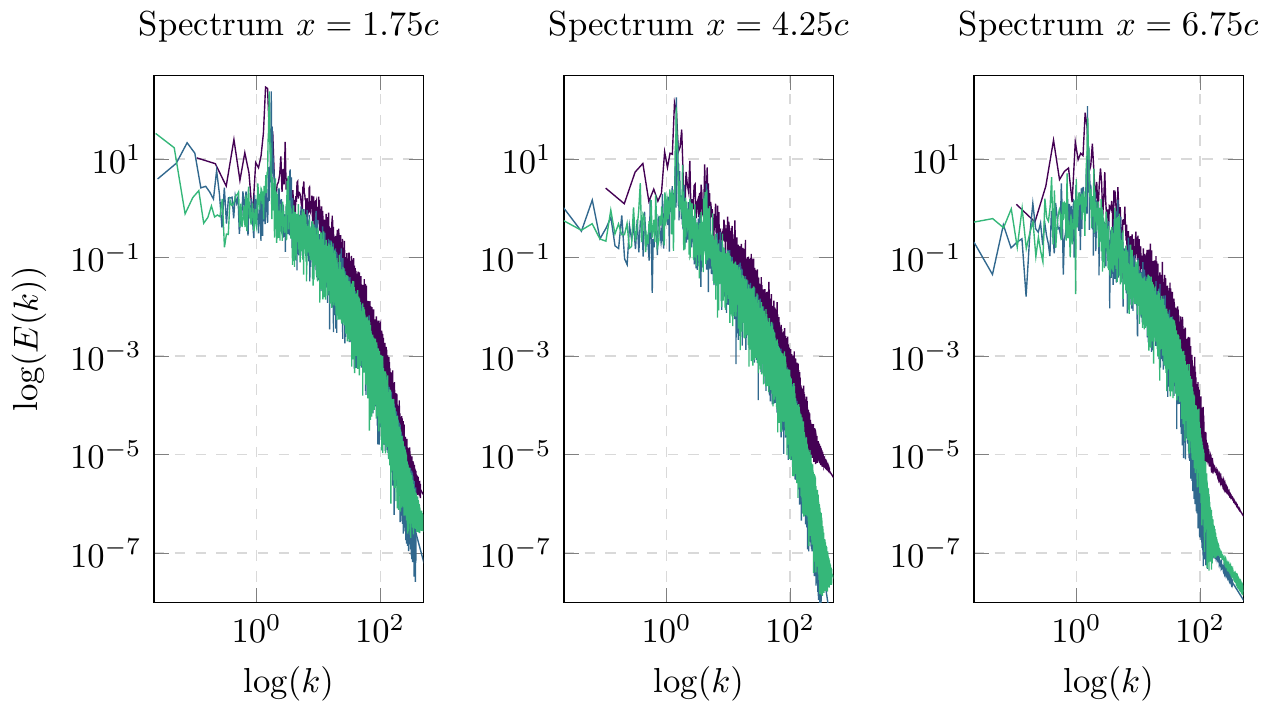}
	\includegraphics{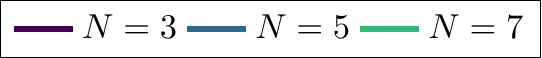}
	\caption{Comparison of the turbulent kinetic energy of different polynomial degrees $N$.}
	\label{fig:FLEXIspectra}
\end{figure}

In \fref{fig:FLEXIspectra} the spectra of the turbulent kinetic energy are shown at three discrete points in the wake of the cylinder. Each figure contains the spectrum for three simulations. Each with different polynomial degree.  

The $N=3$ spectrum in \fref{fig:FLEXIspectra} shows a deviation from the $N=5$ and $N=7$ curves at all evaluation locations. Thus, we can assume that the resolution for $N=3$ is not sufficient for a DNS and does not yield enough dissipation. Since the turbulent kinetic energy spectra for $N=5$ and $N=7$ coincide, we can assume that we are converged at this resolution and thus $N=5$ is sufficient for running a DNS. The expected Strouhal frequency of the cylinder is clearly visible as a distinct peak in the spectrum \cite{Parnaudeau:2008} for all polynomial degrees. 

The simulation with $N=7$ (same amount of degrees of freedom as TAU mesh) is too fine for a typical LES/DNS since the mesh was originally created to be suitable for a hybrid RANS/DNS and a FV code. Evaluating the viscous wall spacing in FLEXI yields $y^+\approx0.01$ which is more than sufficient, even for a DNS. This however underlines the benefits of a high-order scheme when resolving turbulent eddies.

As already shown in \fref{fig:comparison_plot} using the same amount of DOFs in FLEXI and in TAU with a high polynomial degree in FLEXI shows the strength of the high-order scheme, since this resolution is sufficient in FLEXI to run a DNS.

\subsection{Interpolation Error}

\begin{figure}
	\centering
	\includegraphics[width=\columnwidth]{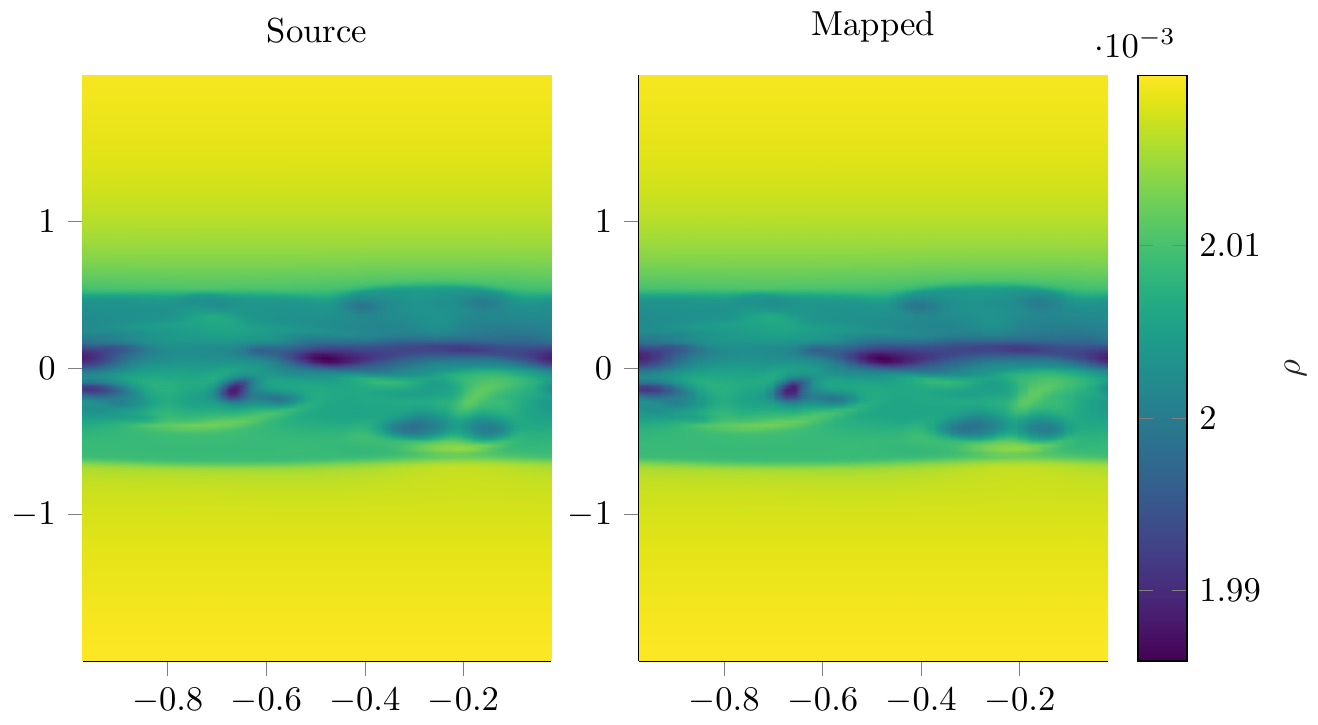}
	\caption{Comparison of the instantaneous flow fields of a cylinder wake state.}
	\label{fig:interpolation_plot}
\end{figure}

Next, we assess the interpolation error resulting from interpolating a wake plane as defined in \fref{fig:cylinder_testcase} onto the FLEXI boundary condition ($\text{grid}_\text{tar}=32\times8$) using the modified Shepard method. We investigate the error for plane $x_{I1}$, which is the one that is located closest to the cylinder. The error is assessed by using the TAU source data as reference data. To get the same point set for TAU and FLEXI we evaluate the polynomials of the mapped FLEXI solution on the TAU solution points. The results are visualized in \fref{fig:interpolation_plot}.

\begin{figure}
	\centering
	\includegraphics[width=\columnwidth]{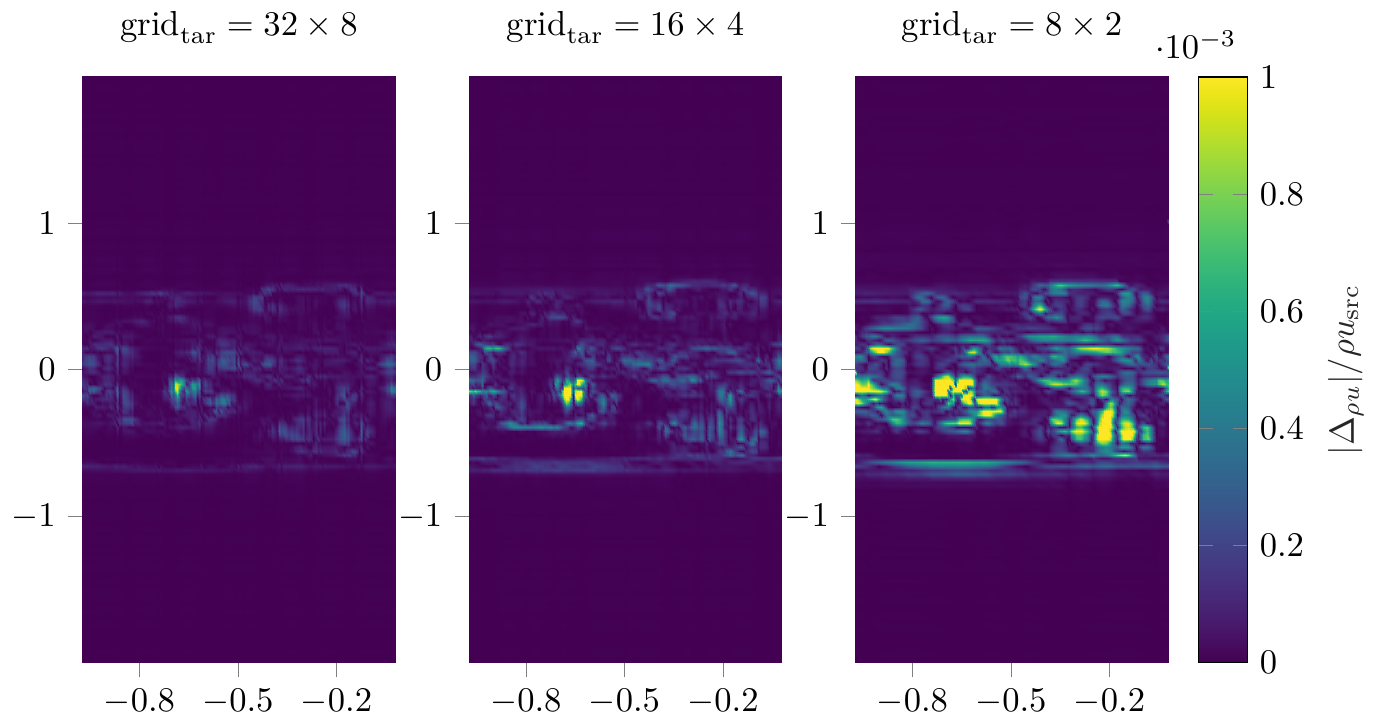}
	\caption{Comparison of the interpolation error of a cylinder wake state.}
	\label{fig:interpolation_error_plot}
\end{figure}

By eye norm the results in \fref{fig:interpolation_plot} look very convincing. The structures of the source data are all represented in the mapped solution. Taking a closer look one can see small overshoots of the mapped solution at the element boundaries. This effect has already been mitigated by using a super-sampling as dealiasing technique. To quantify the error we look at the difference between the source and the mapped data visualized in \fref{fig:interpolation_error_plot}. 

Thus, we evaluate the error of the density for three resolutions $\text{grid}_\text{tar}=32\times8$, $\text{grid}_\text{tar}=16\times4$ and $\text{grid}_\text{tar}=8\times2$. The plots show the relative deviation of the interpolated target data based on the source data. The density plot confirms the observations we made in \fref{fig:interpolation_plot} and shows a very small error in the whole domain. The error increases as espected for coarser resolutions. Calculating the $L_2$-errors for all three meshes yields $L_2\text{-error}(32\times8)$ = \num{5.89E-07}, $L_2\text{-error}(16\times4)$ = \num{3.03E-07} and $L_2\text{-error}(8\times2)$ = \num{9,65E-08} yielding an convergence rate of $EOC=1.31$ which is in line with our findings from \fref{fig:convergence_target} for Shepard's method. Especially in the part containing eddies in the middle of the interface planes we see large errors at the eddy boundaries.

\begin{table}[]
	\centering
	\caption{Minimum, maximum and integral mean values of the primitive variables.}
	\label{tab:interpolation_validation}
	\resizebox{\columnwidth}{!}{\begin{tabular}{cclllll}
		\hline
		& \multicolumn{1}{l}{} & \multicolumn{1}{c}{$\rho$} & \multicolumn{1}{c}{$u$} & \multicolumn{1}{c}{$v$} & \multicolumn{1}{c}{$w$} & \multicolumn{1}{c}{$p$} \\ \hline
		\multirow{2}{*}{Mean} & Source & 2.014E-03 & 2.806E+01  & 2.001E-01  & -6.915E-01 & 1.578E+02 \\
		& Mapped & 2.014E-03 & 2.803E+01  & 2.013E-01  & -6.888E-01 & 1.578E+02 \\
		\multirow{2}{*}{Min.}  & Source & 1.986E-03 & -2.975E+01 & -3.402E+01 & -2.510E+01 & 1.553E+02 \\
		& Mapped & 1.986E-03 & -2.978E+01 & -3.333E+01 & -2.508E+01 & 1.553E+02 \\
		\multirow{2}{*}{Max.}  & Source & 2.020E-03 & 4.800E+01  & 3.178E+01  & 2.967E+01  & 1.583E+02 \\
		& Mapped & 2.020E-03 & 4.787E+01  & 3.153E+01  & 2.932E+01  & 1.583E+02 \\ \cline{1-7} 
	\end{tabular}
	}
\end{table}

In \tref{tab:interpolation_validation} we can see the minimal and maximal values of the primitive variables for source and mapped data. Additionally, the integral mean value is listed. One can note that the mapping yields very good results for density and pressure. In contrast especially for the velocities we see small deviations from source data. This error is based on the fact that the interpolation is not conservative. This gets especially pronounced for the velocity components due to changing signs. By applying the conversion of primitive to conservative variables after interpolation we ensure that - despite the interpolation not being conservative - we get consistent conservative variables.

Hence, due to flexibility of the scheme and the generally very small effect on the mapped results we can neglect the effects of the non-conservativity (\cf \tref{tab:interpolation_validation}) and directly use the mapped plane as an inflow condition.

\subsection{Influence of the Sampling Rate}

Now we assess the effect of the sampling rate of the source data on the quality of the solution in the target domain, which is a very important user parameter that has to be considered when creating a coupled simulation. We do so by investigating the effect on the contribution of the incoming turbulence on the turbulent kinetic energy.

\begin{figure}[!htb]
	\centering
	\includegraphics[width=\columnwidth]{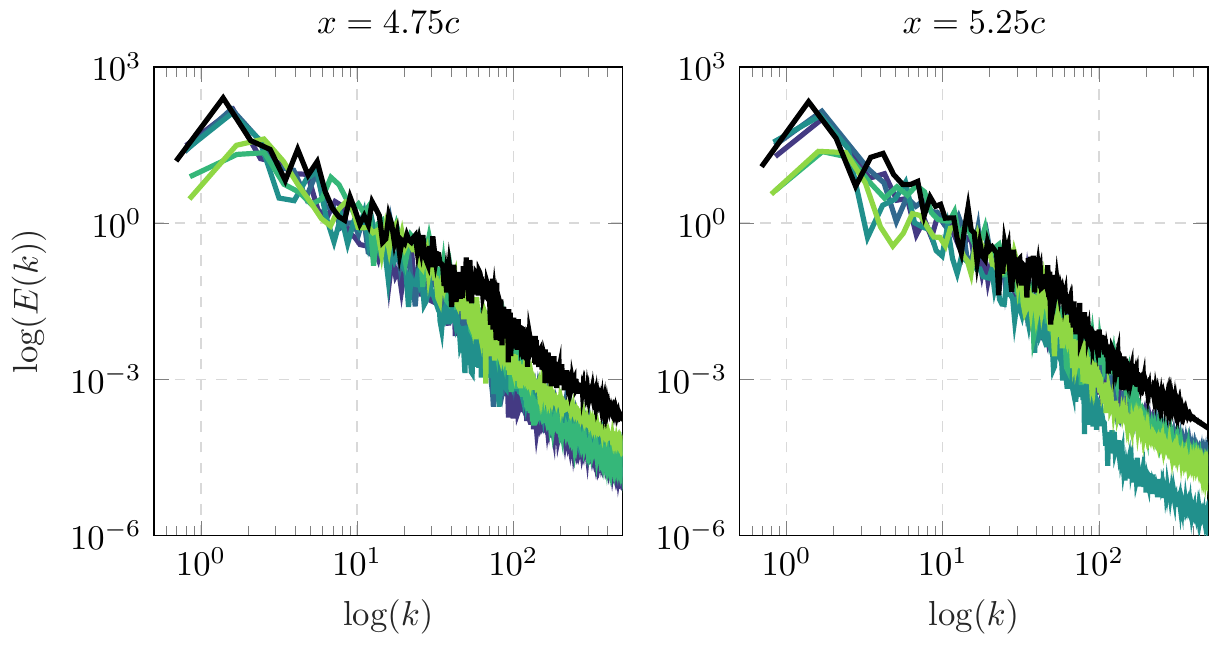}
	\includegraphics[width=\columnwidth]{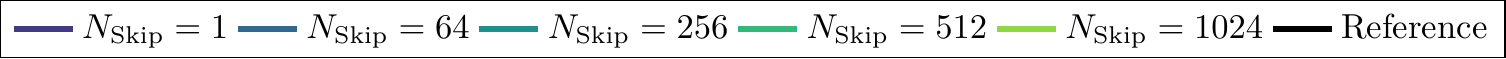}
	\caption{Turbulent kinetic energy at two distinct probe points in the wake of the cylinder with varying sampling rate.}
	\label{fig:sampling_rate_spectrum}
\end{figure}

In \fref{fig:sampling_rate_spectrum} the turbulent kinetic energy spectra at two distinct probe points are visualized. The different colors depict different temporal sampling. With $N_\text{Skip}$ we mean how many TAU snapshots are skipped in time. $N_\text{Skip} = 1$ means that every temporal snapshot is used. The physical TAU sampling rate is $\sim150$ snapshots per characteristic time. The characteristic time is defined as the time it takes the fluid to cover the distance of the diameter of the cylinder. For $N_\text{Skip} = 2$ we only use every second snapshot. The lighter the color gets the fewer snapshots are used to recover the TAU solution in FLEXI. 

\fref{fig:sampling_rate_spectrum} shows that the results are heavily dependent on the sampling rate. This seams reasonable since the sampling rate determines which structures are mapped via the instantaneous boundary condition. According to the Nyquist criterion there is a value for $N_\text{Skip}$ for which the solution is not represented anymore. In this case for $N_\text{Skip}\geq512$ we no longer see agreement with the reference solution. For smaller $N_\text{Skip}$ there is better agreement with the black reference solution. Hence, two major observations can be made. First, for high $N_\text{Skip}$ the major flow structures can not be recovered and even the Strouhal frequency is not represented correctly. Additionally, after some development in the target domain at $x=5.25c$, we can see the there is a lot of disagreement even for low $k$. Second, we can observe that the energy does not adapt and we loose energy in high modes for large $N_\text{Skip}$.

From these observations we can conclude that the sampling frequency is dependent on the structures that have to be mapped to the new domain. Thus, we define a measure to quantify the ``eddy size - sampling rate'' relation which is closely related to the underlying spatial discretization scheme. From literature (\eg \cite{Flad:2020,Gassner:2018}) we know that there is a similar criteria for spatial discretization, which uses the parameter numbers-per-wavelength $n_\text{PPW}$ to quantify the property of a spatial discretization scheme in resolving multi-scale structures. For DGSEM it is known that $n_\text{PPW,DGSEM}\approx4$.

In this case we take two sizes as reference. First, according to \cite{Pope:2015} the large structures are of the size of the cylinder which corresponds to $L=1c$. From the simulation setup and the properties of the DG scheme we estimate the smallest structures according to 
\begin{equation}
	l=\frac{L_\text{domain}}{\text{\#DOF}\cdot n_\text{PPW,DGSEM}}\approx0.06c.
\end{equation}
with $L_\text{domain}$ denoting the size of the domain and $\text{\#DOF}$ the number of DOFs used to discretize the domain.
Taking $u_\infty$ into account, we get an approximation for how long it takes an eddy to be advected over the interface plane, assuming Taylor's hypothesis \cite{Moin:2009}. Taking the sampling frequency into account we can estimate that for the smallest structures we need $N_\text{Skip}\approx4$ and for the large structures $N_\text{Skip}\approx64$ is sufficient. This behavior for $L=c$ is also underlined in \fref{fig:sampling_rate_spectrum}. Only using every \nth{64} sample $N_\text{Skip}=64$ still provides us with the main structures and correct amplitudes, while $N_\text{Skip}>64$ shows signs of underresolution. Using these information we can approximate a criterion on how many points we need per structure/eddy that has to be transported over the interface. It turns out that for both large and small eddies we need approximately $2.3$ samples per eddy. As one would expect, we can conclude that spatial and temporal discretization requirements are similar for the interface. 

We repeated this evaluation for both interface planes $x_\text{I1}$ and $x_\text{I2}$. Both showed qualitatively identical results.

 \section{Summary}\label{summary}

In this work we introduced a method to generate an instantaneous boundary condition relying on a precursor simulation. We presented the numerical methods necessary to handle differences in spatial and temporal discretization via interpolation as well as validated the scheme for simple test cases and a more complex cylinder wake.

We have shown how to generate numerically stable inflow and initial conditions with the methods described in this paper that are universally applicable also to other solvers than TAU and even experimental data. 

The requirements regarding sampling rate are similar to those of the spatial discretization and thus need approximately four sampling points per wavelength, depending on the temporal interpolation scheme used.

We implemented several mapping techniques and showed the differences in interpolation quality and additionally demonstrated their capabilities of reconstructing scattered source data. In addition we utilized super-sampling of the interpolation to increase the overall accuracy and to mitigate the errors due to aliasing and numerical incompatibilities.

In terms of spatial resolution difference at the interface we observed that increasing the resolution of the source data never posed a problem. However, coarsening the data too much can produce large aliasing errors which cause trouble for the high-order scheme. Thus, we recommend using a similar resolution on both sides of the interface.

The introduced interface now has to be applied to more complex scenarios. The tool-chain introduced in this paper is already designed to handle these kind of challenging simulations.
 
\backmatter

\bmhead{Acknowledgments}

The authors gratefully acknowledge the Deutsche Forschungsgemeinschaft DFG (German Research Foundation) for funding this work in the framework of the research unit FOR2895. We also thank the Gauss Centre for Supercomputing e.V. (\url{www.gauss-centre.eu}) for funding this project (GCS-lesdg) by providing computing time on the GCS Supercomputer HAWK at Höchstleistungsrechenzentrum Stuttgart (\url{www.hlrs.de}).

\end{document}